\documentclass[pre,aps,twocolumn,superscriptaddress,floatfix]{revtex4}

\usepackage{mathtools}
\usepackage[usenames,dvipsnames]{xcolor}
\usepackage{amsmath}
\usepackage{amssymb,mathrsfs}
\usepackage{graphicx}
\usepackage[colorlinks=true]{hyperref}
\usepackage{soul}
\usepackage{xcolor}

\usepackage[utf8]{inputenc}
\usepackage[english]{babel}

\newcommand{\om}{\omega}
\newcommand{\la}{\lambda}
\newcommand{\ga}{\gamma}

\newcommand{\vphi}{\varphi}

\newcommand{\prt}{\partial}

\newcommand {\sn} {\mathrm{s n}}

\newcommand {\e} {\mathrm{e}}

\begin{document}

\title{Modulation theory for the sine-Gordon equation
}

\author{A. M. Kamchatnov}  
\affiliation{Institute of Spectroscopy,
Russian Academy of Sciences, Troitsk, Moscow, 108840, Russia}
%  \affiliation{Moscow Institute of
%  Physics and Technology, Institutsky lane 9, Dolgoprudny, Moscow
%  region, 141700, Russia}

\begin{abstract}
We give the solution of the Whitham modulation equations for envelopes of
pulses evolving according to the sine-Gordon equation.
The Whitham equations are interpreted as the equations of 
relativistic hydrodynamics and their solving is reduced by the hodograph 
method to solving a linear partial differential equation. We describe the class of 
solutions of this equation with separation of variables and illustrate the theory 
by an example of the nonlinear wave packet evolution accompanied by its
shrinking and decrease of the number of oscillations in the Whitham
nonlinear region.
\end{abstract}


%05.45.Yv Solitons
%47.35.Fg Solitary waves
% Nonlinear dynamics and chaos
%47.60.+i Flows in ducts, channels, nozzles, and conduits

\maketitle

\section{Introduction}

The phenomenon of modulation instability was discovered independently in several 
physical contexts: as self-focusing of intensive and wide light beams propagating
through a nonlinear medium \cite{askaryan-62,askaryan-73,cgt-64}, as formation of blobs
in a gas of Langmuir plasmons \cite{vr-64,zakharov-72}, as self-contraction of wave
packets in optics \cite{ostrovsky-66} and on deep water \cite{bf-67,zakharov-68}.
If the initial distributions of physical parameters of a wave are modulated by
smooth enough functions, then dynamics of modulations is governed in the main
approximation by hydrodynamic type equations which can be solved by the methods of
the compressible fluid dynamics. First examples of such an approach were given in
the theory of self-focusing where the evolution of light beams is described by the
focusing nonlinear Schr\"{o}dinger (NLS) equation so that the modulation parameters
are the light intensity and the transverse wave number of the light wave. They obey
the geometric optics equations equivalent to the hydrodynamic equations for 
``inverted shallow water'' waves. In this case, V.~I.~Talanov found the solution
for evolution of a light beam with a parabolic initial intensity profile
\cite{talanov-65}, and S.~A.~Akhmanov, A.~P.~Sukhorukov, and R.~V.~Khokhlov \cite{ask-66} 
for beams with initial profiles of the type $\cosh^{-2}(x)$. In the papers 
\cite{lighthill-1965,hayes-73} a similar approach was formulated without derivation
of the NLS equation, however also in the approximation of the moderate wave intensity,
and the resulting modulation hydrodynamic system was reduced by the hodograph method
to the linear equation of elliptic type. Some examples of its solutions for
self-focusing of beams and self-contraction of pulses were given in Ref.~\cite{gs-70} 
and many other examples can be found in the book \cite{zt-91}.

Naturally, these solutions assuming smooth modulation of a wave are only correct up to
the self-focusing moment. Besides that, they are unstable with respect to small
perturbations violating smoothness of the wave profile. Already in 
Refs.~\cite{karpman-67,kk-68} it was noticed that a localized initial perturbation of a
uniform plane leads in the NLS equation theory to formation of an expanding strongly
modulated wave structure. Application to its evolution of the Whitham modulation
theory \cite{whitham-65,whitham-77,fl-86,pavlov-87} showed \cite{kamch-92,ehkk-93,bk-94}, 
that the front of instability propagates with the minimal group velocity corresponding
to the real branch of the dispersion relation and this result was confirmed in 
Ref.~\cite{bm-16} by the inverse scattering transform method. Generalization \cite{ks-21} 
of this theory to non-uniform evolving distributions permitted one to find the laws of
propagation of the boundaries of the instability region for arbitrary initial profiles
of the unstable pulse.

The derivation of the results described above is based on the assumption that in the main
approximation the wave is linear and the nonlinearity introduces only a small 
intensity-dependent correction into the dispersion relation. In geometrical optics 
approximation this is equivalent to the dependence of the refraction index on the intensity,
and then the geometrical optics equations are equivalent to the hydrodynamic equations for
the ``inverted shallow water'' which simplicity allows one to use the well-developed
methods of gas dynamics. However, the situation is different in case of the well-known
nonlinear Klein-Gordon equation
\begin{equation}\label{eq1}
  \vphi_{tt}-\vphi_{xx}+U'(\vphi)=0,\quad U'=\frac{dU}{d\vphi},\quad U'(0)=0,
\end{equation}
which has numerous applications to various physical problems, especially in a particular
case of the sine-Gordon equation with $U'(\vphi)=\sin\vphi$ (see, e.g., \cite{scott-07,cmkw} 
and references therein). This equation has traveling wave solutions $\vphi=\vphi(A,kx-\om t)$
in which the dependence of the frequency $\om$ on the wave amplitude $A$ cannon be considered 
as small anymore. In a modulated wave its amplitude $A$ and phase velocity $V=\om/k$ become
slow functions of the space coordinate $x$ and time $t$ which change little in one wave length
and one period. The corresponding modulation equations for dynamics of $A$ and $V$ were obtained
by Whitham \cite{whitham-65,whitham-77}. In Ref.~\cite{maslov-69} similar equations for the
quasi-classical asymptotic solutions of Eq.~(\ref{eq1}) were derived by the Bogoliubov-Mitropolsky 
method and it was indicated that these equations are equivalent to equations of relativistic
hydrodynamics (see also the review article \cite{dm-80}). It is essential that in the most
important case of the sine-Gordon equation this dynamics is modulationally unstable, however,
because of complexity of the equation of state related with the dispersion relation
$\om=\om(k,A)$ for waves in the ``effective relativistic matter'' and of complexity of the
equations of the relativistic hydrodynamics, this theory has not been applied yet to any
concrete problems about evolution of wave packets.

The aim of this paper is to develop the method of solving the Whitham modulation equations for
one-phase traveling waves whose evolution obeys the sine-Gordon equation. Earlier the equations
of relativistic hydrodynamics were studied in the theory of multiple particle creation in
ultra-relativistic collisions of nuclei and nucleons \cite{landau-53,khal-54,nk-76,kamch-19}. 
We will show that the methods used in these papers in case of extremely simple equation of state
$p=e/3$ ($p$ is pressure, $e$ is energy density) can be modified to become applicable
to much more complicated equation of state corresponding to the sine-Gordon equation. As a result,
we shall derive a linear partial differential equation whose solutions define solutions of the
relativistic hydrodynamics equations in the hodograph method. We shall describe a class of solutions
of this equation with separation of variables and illustrate the theory by an example of the
solution for the self-contracting wave packet when its evolution is accompanied by decrease of
the number of waves in the region of nonlinear oscillations.

\section{Whitham equations}

First, we reproduce here the main relations of the Whitham modulation theory 
\cite{whitham-65,whitham-77} for the nonlinear Klein-Gordon equation (\ref{eq1}). It is easy to
see that this equation has traveling wave solutions $\vphi=\vphi(\xi)$, $\xi=x-Vt$, where $\vphi(\xi)$ 
is defined implicitly by the equation 
\begin{equation}\label{eq2}
  \xi-\xi_0=\sqrt{\frac{V^2-1}{2}}\int_{\vphi_0}^{\vphi}\frac{d\vphi}{\sqrt{A-U(\vphi)}},
\end{equation}
so that $V$ and the integration constant $A$ are parameters, $\vphi(\xi_0)=\vphi_0$, and the variable
$\vphi$ oscillates between two roots of the equation $A-U(\vphi)=0$ in the positivity interval of
this expression. Following Whitham \cite{whitham-65,whitham-77}, we define the function
\begin{equation}\label{eq3}
\begin{split}
  W(V,A)&=\sqrt{2(V^2-1)}\oint \sqrt{A-U(\vphi)}\,d\vphi\\
  &\equiv\sqrt{V^2-1}\cdot G(A),
  \end{split}
\end{equation}
where the integration is taken along a contour around this positivity interval. Then the wavelength 
of the above solution is given by the expression
\begin{equation}\label{eq4}
  L=\frac{\prt W}{\prt A}=\sqrt{V^2-1}\cdot G'(A).
\end{equation}
We define the wave number as $k=1/L$, so that
$k^2(V^2-1)=(G')^{-2}$, and obtain for the frequency $\om=kV$ the dispersion relation
\begin{equation}\label{eq5}
  \om^2=k^2+(G'(A))^{-2},
\end{equation}
which depends essentially on the amplitude $A$. The group velocity is defined as 
\begin{equation}\label{eq6}
  v=\left(\frac{\prt\om}{\prt k}\right)_A=\frac{k}{\om}=\frac{1}{V}.
\end{equation}

In a modulated wave the parameters $V$ and $A$ become slow functions of $x$ and $t$, and their evolution
obeys the Whitham modulation equations \cite{whitham-65,whitham-77} which in our notation can be
written in the form
\begin{equation}\label{eq7}
  \begin{split}
  & \left(\frac{kV}{V^2-1}+A\right)_t+\left(\frac{kVW}{V^2-1}\right)_x=0,\\
  & \left(\frac{kVW}{V^2-1}\right)_t+\left(\frac{kV^2W}{V^2-1}-A\right)_x=0.
  \end{split}
\end{equation}
For relativistic interpretation of these equations it is convenient to exclude $W$ and $V$ with
the use of Eqs.~(\ref{eq3}), (\ref{eq6}) and then we get
\begin{equation}\label{eq8}
  \begin{split}
  & \left(\frac{G/G'}{1-v^2}+A-\frac{G}{G'}\right)_t+\left(\frac{(G/G')v}{1-v^2}\right)_x=0,\\
  & \left(\frac{(G/G')v}{1-v^2}\right)_t+\left(\frac{(G/G')v^2}{1-v^2}-A+\frac{G}{G'}\right)_x=0.
  \end{split}
\end{equation}
Linearization of these equations with respect to small deviations from constant values
$A$ and $v$ yields the characteristic velocities
\begin{equation}\label{eq9}
  v_{\pm}=\frac{v\pm c}{1\pm vc},
\end{equation}
where $c$ is defined by the expression 
\begin{equation}\label{eq10}
  c^2=-\frac{GG^{\prime\prime}}{(G')^2}.
\end{equation}
If its right-hand side is positive, then formulas (\ref{eq9}) have simple physical sense: they
give velocities of propagation of a sound signal with the speed $c$ upstream or downstream the
flow of a ``fluid'' moving with velocity $v$, so that in the laboratory reference frame the
velocity of the signal is equal to the relativistic sum of these two velocities.

Equations (\ref{eq8}) can be transformed to the diagonal form 
\begin{equation}\label{eq11}
  \frac{\prt r_{\pm}}{\prt t}+v_{\pm}\frac{\prt r_{\pm}}{\prt x}=0
\end{equation}
by introduction of the Riemann invariants 
\begin{equation}\label{eq12}
  r_{\pm}=\int^v\frac{dv}{1-v^2}\pm\int^A\frac{cG'}{G}\,dA,
\end{equation}
and this simplifies essentially solving concrete problems in case of real sound velocity
$c$. However, we are interested in the opposite case of imaginary ``sound velocity'' $c$,
so let us discuss in some detail the properties of relativistic hydrodynamics for the
sine-Gordon model.

\section{Relativistic hydrodynamics for the nonlinear Klein-Gordon equation}

Already Whitham in his fundamental paper \cite{whitham-65} remarked that his averaging procedure
of conservation laws over fast oscillations is analogous to transition from microscopic
description of medium's motion in statistical mechanics to the averaged hydrodynamic description
which is correct under condition of smallness of gradients of physical variables.
In our case these parameters are $A$ and $V$, Eq.~(\ref{eq1}) is relativistically invariant,
so it is natural to expect that after averaging the conservation laws we must arrive at the
equations of relativistic hydrodynamics. In framework of the asymptotic WKB method this was
shown in Ref.~\cite{maslov-69}, but for transition from Eqs.~(\ref{eq8}) to equations of
relativistic hydrodynamics it is convenient to start from the equations of the energy-momentum
conservation law in a relativistic flow (see\cite{LL6}),
\begin{equation}\label{eq13}
  \frac{\prt T^{00}}{\prt t}+\frac{\prt T^{10}}{\prt x}=0,\quad
  \frac{\prt T^{10}}{\prt t}+\frac{\prt T^{11}}{\prt x}=0,
\end{equation}
where
\begin{equation}\label{eq14}
  T^{ij}=wu^iu^j-pg^{ij},\quad i,j=1,2,
\end{equation}
is the energy-momentum tensor in two-dimensional Minkowski space with the metric tensor 
\begin{equation}\label{eq15}
  g^{ij}=\left(
           \begin{array}{cc}
             1 & 0 \\
             0 & -1 \\
           \end{array}
         \right),
\end{equation}
$w=e+p$ is the enthalpy density, $e$ is the energy density, $p$ is the pressure, and
$u^i$ is a two-dimensional vector of ``4-velocity''. To identify the Whitham equations (\ref{eq8}) 
with Eqs.~(\ref{eq13}), we introduce in a usual way the 4-vector  $u^i$,
\begin{equation}\label{eq16}
  u^0=\frac{1}{\sqrt{1-v^2}},\qquad u^1=\frac{v}{\sqrt{1-v^2}},
\end{equation}
and then it is easy to see that equations (\ref{eq8}) and (\ref{eq13}) coincide with each other if
\begin{equation}\label{eq17}
  \begin{split}
  & T^{00}=\frac{w}{1-v^2}-p=\frac{G/(2G')}{1-v^2}+\frac{A}2-\frac{G}{2G'},\\
  & T^{10}=T^{01}=\frac{wv}{1-v^2}=\frac{(G/(2G'))v}{1-v^2},\\
  & T^{11}=\frac{wv^2}{1-v^2}+p=\frac{G/(2G')}{1-v^2}-\frac{A}2+\frac{G}{2G'},
  \end{split}
\end{equation}
where we divided Eqs.~(\ref{eq8}) by 2 for further convenience. Consequently we get the 
relationships of the wave amplitude $A$ with thermodynamical functions of effective matter whose
dynamics obeys equations (\ref{eq13}),
\begin{equation}\label{eq18}
  e=\frac{A}{2},\quad p=\frac{G}{2G'}-\frac{A}2,\quad w=e+p=\frac{G}{2G'}.
\end{equation}
It is important that the pressure $p$ only depends on the energy density $e$. This means that the 
mass of particles in the effective matter is negligibly small.
The expression (\ref{eq10}) for the sound velocity can be written in the standard form 
\begin{equation}\label{eq19}
  c^2=\frac{dp}{de}.
\end{equation}
The temperature $T$ and the entropy density $\sigma$ of the effective matter can be defined in 
the following way. The chemical potential of a gas of massless particles equals to zero, so
the enthalpy is given by the formula $w=T\sigma$, consequently the relationship 
$dw=Td\sigma+dp=d(T\sigma)$ gives $dp=\sigma dT$
(see \cite{LL5}). Hence, the squared sound velocity can be written in two forms: from (\ref{eq10}) 
and (\ref{eq18}) we get
$$
c^2=-w\frac{d^2G/de^2}{dG/de},
$$
whereas Eq.~(\ref{eq19}) with account of $dp=\sigma dT=(w/T)dT$ yields
$$
c^2=\frac{w}{T}\frac{dT}{de}.
$$
Comparison of these two expressions gives the formulas
\begin{equation}\label{eq20}
  T=\gamma\cdot\left(\frac{dG}{de}\right)^{-1},\qquad \sigma=\frac1{\gamma}\cdot G(e)
\end{equation}
with the same numerical factor $\gamma$ in both formulas. Then we obtain the relation 
\begin{equation}\label{eq20b}
  d\sigma=\frac1{\ga}\frac{dG}{de}de=\frac{de}{T},
\end{equation}
which agrees with the standard thermodynamical definition of the entropy.

The hydrodynamic equations get especially simple form in the variables $T,\sigma, u^i$. 
We notice that the formulas for the wave vector $k=1/L$ and the frequency $\om=kV=k/v$ 
are transformed to
\begin{equation}\label{eq21}
\begin{split}
  & k=2u^1\left(\frac{dG}{de}\right)^{-1}=\frac2{\ga}u^1T,\\
  & \om=k\frac{u^0}{u^1}=\frac2{\ga}u^0T.
  \end{split}
\end{equation}
Consequently the conservation of the number of waves law, which follows from Eqs.~(\ref{eq7})
(see \cite{whitham-65,whitham-77}) 
\begin{equation}\label{eq22}
  k_t+\om_x=0,
\end{equation}
transforms to
\begin{equation}\label{eq23}
  (u^1T)_t+(u^0T)_x=0.
\end{equation}
It is worth noticing that this relation was obtained by I.~M.~Khalatnikov \cite{khal-54} 
from Eqs.~(\ref{eq13}) for any one-dimensional relativistic flow what proves that it is
potential. In Whitham's theory Eq.~(\ref{eq22}) follows from definition of the wave vector
and the frequency as the phase derivatives, $k=\theta_x,\om=-\theta_t$. We see that both
pictures, Whitham's modulational and relativistic hydrodynamical ones, agree mathematically 
and differ only in notation and physical meaning of variables. 

One more equation we obtain from the formula 
\begin{equation}\label{eq24}
  \frac{\prt(wu^i)}{\prt x^i}-u^i\frac{\prt p}{\prt x^i}=0,
\end{equation}
which is a consequence of Eqs.~(\ref{eq13}), (\ref{eq14}) (see Eq.~(134.5) and problem 2 
in Ref.~\cite{LL6}, \S134). Substitution of $w=T\sigma$, $dp=\sigma dT$ gives at once
the equation
\begin{equation}\label{eq25}
  \frac{\prt(\sigma u^0)}{\prt t}+\frac{\prt(\sigma u^1)}{\prt x}=0,
\end{equation}
which means conservation of entropy, i.e. the flow is adiabatical.

Let us specify these relation for the sine-Gordon equation when we have in Eq.~(\ref{eq1})
\begin{equation}\label{eq26}
  U'(\vphi)=\sin\vphi,\qquad U(\vphi)=1-\cos\vphi.
\end{equation}
Then the integral in Eq.~(\ref{eq2}) reduces to the elliptical one of the 1st kind and its
inversion yields the periodic solution in explicit form 
\begin{equation}\label{eq26a}
\begin{split}
  \vphi&=2\arcsin\left[\sqrt{e}\,\,\sn\left(\frac{\xi-\xi_0}{\sqrt{V^2-1}},e\right)\right]=\\
  &=2\arcsin\left[\sqrt{e}\,\,\sn\left(\frac{v(x-x_0)-t}{\sqrt{1-v^2}},e\right)\right].
  \end{split}
\end{equation}
In the limit $e\to1$ this solution converts into the known kink solution of the sine-Gordon equation 
\begin{equation}\label{eq26b}
  \vphi=2\arcsin\left[\sqrt{e}\,\,\tanh\left(\frac{v(x-x_0)-t}{\sqrt{1-v^2}},e\right)\right],
\end{equation}
so that $\vphi=-\pi$ as $x\to-\infty$ and $\vphi=\pi$ as $x\to+\infty$. The integral in 
Eq.~(\ref{eq3}) can be calculated with the help of substitution 
$\sin(\vphi/2)=\sqrt{e}\sin\psi$,
\begin{equation}\label{eq27}
\begin{split}
  G(e)&=2\sqrt{2}\int_{-\vphi_0}^{\vphi_0}\sqrt{2e-1+\cos\vphi}\,d\vphi=\\
  &=16\{E(e)-(1-e)K(e)\},
  \end{split}
\end{equation}
where $K(e),E(e)$ are the complete elliptic integrals of the 1st and 2nd kind, respectively,
defined here according to the handbook \cite{as-79} ($\pm\vphi_0$ are the roots of the integrand
function). We choose for the factor $\ga$ in Eqs.~(\ref{eq20}) the value $\ga=8$ and then the formulas 
\begin{equation}\label{eq28}
  \frac{dK}{de}=\frac{E-(1-e)K}{2e(1-e)},\qquad \frac{dE}{de}=\frac{E-K}{2e}
\end{equation}
for derivatives of the elliptic integrals yield the expressions for the thermodynamic functions
\begin{equation}\label{eq29}
  T=\frac1{K},\quad\sigma=4e(1-e)\frac{dK}{de}=2[E-(1-e)K].
\end{equation}
Equation of state $p=p(e)$ has the form
\begin{equation}\label{eq30}
  p=2\left(\frac{E(e)}{K(e)}-1\right)+e.
\end{equation}
The squared sound velocity equals to
\begin{equation}\label{eq31}
  c^2=-4e(1-e)\left(\frac1K\frac{dK}{de}\right)^2
\end{equation}
and it is negative in the region $0\leq e<1$ of existence of periodic solutions, that is they are
modulationally unstable. The corresponding Riemann invariants (\ref{eq12}) are complex:
\begin{equation}\label{eq31b}
  r_{\pm}=\frac12\ln\frac{1+v}{1-v}\pm i\arcsin\sqrt{e}.
\end{equation}
In spite of that, as in the NLS equation theory, evolution of the modulation parameters obeys
the hydrodynamic equations (\ref{eq23}), (\ref{eq25}) and their solution can be obtained by means
of the hodograph transform.  

\section{Hodograph transform}

To perform the hodograph transform, we introduce `rapidity' $y$ instead of velocity $v$ 
according to the relations
\begin{equation}\label{eq32}
  u^0=\cosh y,\qquad u^1=\sinh y,\qquad v=\tanh y,
\end{equation}
and pass to the `light-cone' variables
\begin{equation}\label{eq33}
x_-=t-x,\qquad x_+=t+x.
\end{equation}
Then Eq.~(\ref{eq23}) takes the form
\begin{equation}\label{eq34}
  \frac{\prt}{\prt x_-}\left(\frac{\e^{-y}}{K(e)}\right)-
  \frac{\prt}{\prt x_+}\left(\frac{\e^{y}}{K(e)}\right)=0,
\end{equation}
and it is satisfied if 
\begin{equation}\label{eq35}
  \frac{\e^{-y}}{K(e)}=\frac{\prt\phi}{\prt x_-},\qquad
  \frac{\e^{y}}{K(e)}=\frac{\prt\phi}{\prt x_+}
\end{equation}
for some potential $\phi=\phi(x_-,x_+)$. Now, following Khalatnikov \cite{khal-54}, we make the
Legendre transform to the potential $\mathcal{W}=\mathcal{W}(e,y)$,
\begin{equation}\label{eq36}
  \mathcal{W}=\phi-K^{-1}\e^yx_--K^{-1}\e^{-y}x_+,
\end{equation}
so that
\begin{equation}\label{eq37}
\begin{split}
  d\mathcal{W}=&-\frac{dK^{-1}}{de}(\e^yx_-+\e^{-y}x_+)de-\\
  &-\frac1K(\e^yx_--\e^{-y}x_+)dy,
  \end{split}
\end{equation}
and, hence,
\begin{equation}\label{eq38}
\begin{split}
  & x_-=-\frac{\e^{-y}}{2}\left(\frac{1}{dK^{-1}/de}\frac{\prt\mathcal{W}}{\prt e}
  +K\frac{\prt\mathcal{W}}{\prt y}\right),\\
  & x_+=-\frac{\e^{y}}{2}\left(\frac{1}{dK^{-1}/de}\frac{\prt\mathcal{W}}{\prt e}
  -K\frac{\prt\mathcal{W}}{\prt y}\right).
  \end{split}
\end{equation}
These formulas correspond to the hodograph transform: if the function $\mathcal{W}=\mathcal{W}(e,y)$
is known, then they give the dependence of $e$ and $y$ on the variables (\ref{eq33}) and,
consequently, on $x$ and $t$.

Equation (\ref{eq25}) after substitutions of (\ref{eq32}), (\ref{eq33}) transforms to
\begin{equation}\label{eq39}
\begin{split}
  &\frac{\prt(\sigma \e^{-y})}{\prt x_-}+\frac{\prt(\sigma \e^y)}{\prt x_+}=\\
 &= \frac{\prt(\sigma \e^{-y},x_+)}{\prt (x_-,x_+)}+\frac{\prt(x_-,\sigma \e^y)}{\prt(x_-, x_+)}=0.
  \end{split}
\end{equation}
Multiplying it by the Jacobian $\prt(x_-,x_+)/\prt(e,y)$, we obtain after simple transformations
with account of Eqs.~(\ref{eq29}) and $d\sigma/de=1/T=K(e)$ the equation for $\mathcal{W}$:
\begin{equation}\label{eq41}
  \frac{\prt}{\prt e}\left[e(1-e)K^2\frac{\prt\mathcal{W}}{\prt e}\right]+
  \frac{K^2}{4}\frac{\prt^2\mathcal{W}}{\prt y^2}=0
\end{equation}
or
\begin{equation}\label{eq42}
  e(1-e)\frac{\prt^2\mathcal{W}}{\prt e^2}+
  \left(\frac{E(e)}{K(e)}-e\right)\frac{\prt\mathcal{W}}{\prt e}
  +\frac14\frac{\prt^2\mathcal{W}}{\prt y^2}=0.
\end{equation}
This equation is of elliptic type and it replaces the Euler-Poisson equation appearing in
applications of the hodograph transform to the gas dynamics equations (see, e.g., \cite{LL6}).
In modulationally unstable dispersionless dynamics of the NLS equation theory the
analogous equation is the two-dimensional Laplace equation written in polar coordinates
(see, e.g., \cite{lighthill-1965,hayes-73,gs-70}). Equation (\ref{eq41}) transforms to this
Laplace equation in the limit of small $e$.

An important class of solutions of Eq.~(\ref{eq42}) is obtained after separation of variables,
\begin{equation}\label{eq43}
  \mathcal{W}(e)=Z(e)Y(y),
\end{equation}
so that
\begin{equation}\label{eq44}
  \frac{d^2Y}{dy^2}=\la^2Y,\qquad Y(y)=\e^{\pm\la y},
\end{equation}
and $Z=Z(e)$ satisfies the equation
\begin{equation}\label{eq45}
  e(1-e)\frac{d^2Z}{d e^2}+ \left(\frac{E(e)}{K(e)}-e\right)\frac{dZ}{d e}
  +\frac{\la^2}4Z=0.
\end{equation}
One can check with the use of formulas (\ref{eq28}) that for $\la^2=1$ its particular solution
is given by $Z=1/K(e)$. Therefore we make in Eq.~(\ref{eq45}) the substitution
\begin{equation}\label{eq46}
  Z=\frac{F(e)}{K(e)}
\end{equation}
and obtain for $F(e)$ the hypergeometric equation 
\begin{equation}\label{eq47}
  e(1-e)\frac{d^2F}{d e^2}+ (1-2e)\frac{dF}{d e}
  +\frac{\la^2-1}4F=0.
\end{equation}
Its solution is the hypergeometric function $F=F((1+\la)/2,(1-\la)/2,1;e)$
(see, e.g., \cite{as-79,ww-63}). The substitution $e=(1-z)/2$ casts it to the Legendre equation 
\begin{equation}\label{eq48}
  (1-z^2)\frac{d^2F}{dz^2}-2z\frac{dF}{dz}+n(n+1)F=0,\quad n=\frac{\la-1}{2},
\end{equation}
which for integer $n=0,1,2,\ldots$ has solutions in the form of Legendre polynomials 
$P_n(z)=P_n(1-2e)$ and Legendre functions of 2nd kind $Q_n(z)=Q_n(1-2e)$. It is clear that any
linear combination of solutions (\ref{eq43}) is also a solution of Eq.~(\ref{eq42}).

Let us illustrate the developed here theory by an example.

\section{Example: self-contraction of a nonlinear wave packet}

\begin{figure}
\includegraphics[width=8cm]{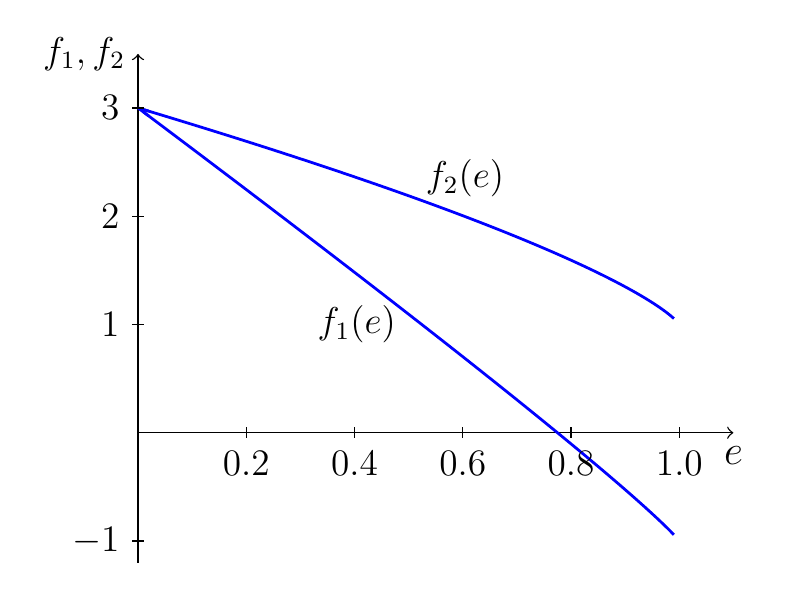}
\caption{Plots of functions $f_1(e)$ and $f_2(e)$ defined by Eqs.~(\ref{eq53}).}
\label{fig1}
\end{figure}

Let us take as an illustrative example the solution with $F=P_1(z)=1-2e$, that is with $n=1$ and $\la=3$:
\begin{equation}\label{eq51}
  \mathcal{W}(e,y)=\frac{1-2e}{K(e)}\e^{3y}.
\end{equation}
Its substitution into Eqs.~(\ref{eq38}) gives
\begin{equation}\label{eq52}
\begin{split}
  &x_-=t-x=-2f_1(e)\e^{2y},\\
  &x_+=t+x=-f_2(e)\e^{4y},
  \end{split}
\end{equation}
where
\begin{equation}\label{eq53}
  \begin{split}
  & f_1(e)=\frac{(1-2e)E(e)-(1-e)(1-3e)K(e)}{E(e)-(1-e)K(e)},\\
  & f_2(e)=-\frac{(1-2e)E(e)-(1-e)K(e)}{E(e)-(1-e)K(e)}.
  \end{split}
\end{equation}
Plots of these functions are shown in Fig.~\ref{fig1}. Their values at $e=0$ and $e=1$ are equal to
\begin{equation}\label{eq54}
\begin{split}
  &f_1(0)=f_2(0)=3,\\
  &f_1(1)=-1,\quad f_2(1)=1,
  \end{split}
\end{equation}
and series expansions in vicinity of $e=0$ have the form ($0<e\ll1$)
\begin{equation}\label{eq55a}
  \begin{split}
  &f_1(e)=3-\frac{15}{4}e-\frac{3}{32}e^2+\ldots,\\
  &f_2(e)=3-\frac{3}{2}e-\frac{3}{16}e^2+\ldots.
  \end{split}
\end{equation}

Formulas (\ref{eq52}) define implicitly the dependence of the variables $e$ and $y$ on $x$ and $t$.
It is convenient to express all variables at fixed value of time $t$ parametrically with $e$
playing the role of the parameter. Indeed, from Eq.~(\ref{eq52}) we find
\begin{equation}\label{eq55}
  \begin{split}
   2t=-2f_1\e^{2y}-f_2\e^{4y},\quad
   2x=2f_1\e^{2y}-f_2\e^{4y},
  \end{split}
\end{equation}
so that the first equation gives
\begin{equation}\label{eq56}
  \e^{2y}\equiv\frac{1+v}{1-v}=\sqrt{\left(\frac{f_1}{f_2}\right)^2-\frac{2t}{f_2}}
  -\frac{f_1}{f_2},
\end{equation}
where we have chosen the positive root due to positivity of the function  $\e^{2y}$. Substitution
of $\e^{2y}$ into the second formula (\ref{eq55}) gives
\begin{equation}\label{eq57}
  x=x(e)=t+2f_1\left(\sqrt{\left(\frac{f_1}{f_2}\right)^2-\frac{2t}{f_2}}
  -\frac{f_1}{f_2}\right).
\end{equation}
This formula defines the dependence of $e$ on $x$ at the fixed moment of time $t$.
The dependence of the velocity $v$ on $e$ we find from Eq.~(\ref{eq56}):
\begin{equation}\label{eq58}
  v=v(e)=\frac{\sqrt{f_1^2-2f_2t}-f_1-f_2}{\sqrt{f_1^2-2f_2t}-f_1+f_2}.
\end{equation}
This formula together with Eq.~(\ref{eq57}) define the distribution of $v$ on $x$ in parametric form.
The obtained here formulas give a particular solution of the Whitham equations about
evolution of a nonlinear wave packet in the sine-Gordon equation theory. Let us discuss its
characteristic features.

\begin{figure}
\includegraphics[width=8cm]{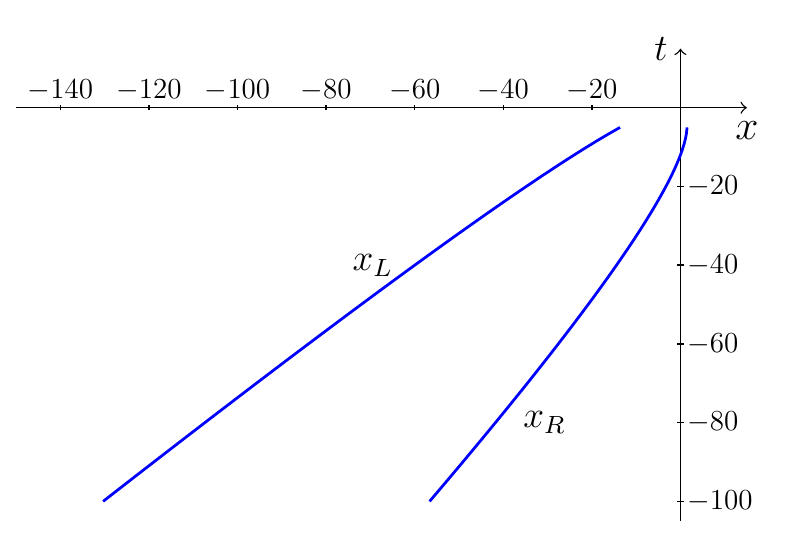}
\caption{Paths of the left $x_L$ and right $x_R$ edges of the nonlinear wave packet 
in the $(x,t)$-plane calculated according to Eqs.~(\ref{eq61}) and (\ref{eq59}), respectively.}
\label{fig2}
\end{figure}

First of all we see that since $f_2$ is positive and both functions $f_1,f_2$ are bounded,
we must have $t<0$, that is the packet was formed at some moment $t<0$ with large enough $|t|$
and after that it evolves with decrease of $|t|$. If we put $e=0$ in Eq.~(\ref{eq57}), we find
with account of (\ref{eq54}) the law of motion of the small-amplitude edge of the packet:
\begin{equation}\label{eq59}
  x_R(t)=t+6\left(\sqrt{1-2t/3}-1\right).
\end{equation}
Its velocity equals to 
\begin{equation}\label{eq60}
  \frac{dx_R}{dt}=1-\frac{2}{\sqrt{1-2t/3}},
\end{equation}
that it coincides with the group velocity (\ref{eq58}) at $e=0$: $dx_R/dt=v(0)$.

The law of motion of the opposite left edge of the packet with $e=1$ is found from Eq.~(\ref{eq57})
by putting the corresponding values from Eq.~(\ref{eq54}):
\begin{equation}\label{eq61}
  x_L(t)=t-2\sqrt{1-2t}-2,
\end{equation}
so that its velocity equal to
\begin{equation}\label{eq62}
  \frac{dx_L}{dt}=1+\frac{2}{\sqrt{1-2t}}.
\end{equation}
The group velocity (\ref{eq58}) at this edge has the value
\begin{equation}\label{eq63}
  v(1)=\frac{\sqrt{1-2t}}{\sqrt{1-2t}+2},
\end{equation}
that is by virtue of Eq.~(\ref{eq6}) this edge velocity (\ref{eq62}) is equal to the phase velocity
of the wave at this point: $dx_L/dt=V=1/v(1)$. This means that at the edge with $e\to1$ we get
a train of kinks (\ref{eq26b}): kink's velocity is given, evidently, by $V=1/v$. The paths of the
edges of the wave packet are shown in Fig.~\ref{fig2} and it is clearly seen that the packet
shrinks with growth of time. 

\begin{figure}
\includegraphics[width=8cm]{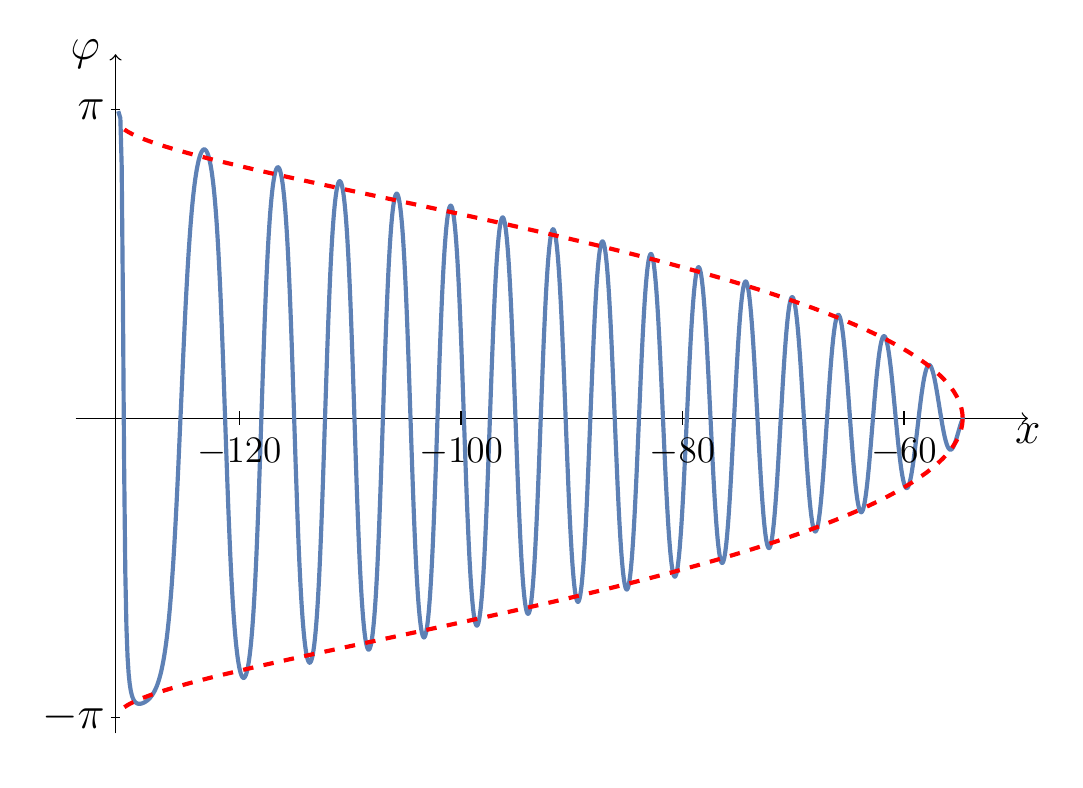}
\caption{Wave's profile in the modulated nonlinear packet $\vphi(x,t)$ at $t=-100$.
}
\label{fig3}
\end{figure}

Substitution of formulas for $x=x(e)$ and $v=v(e)$ from (\ref{eq57}), (\ref{eq58}) to the periodic
solution (\ref{eq26a}) yields together with Eq.~(\ref{eq57}) the profile of the modulated wave packet
at the moment $t$ (naturally, we have neglected here an additional slow phase shift along the wave 
structure; see, e.g., \cite{dm-11}), and the envelopes of this profile are defined, evidently,
by the formula
\begin{equation}\label{eq64}
  a=\pm2\arcsin\sqrt{e(x)}.
\end{equation}
A typical profile is shown in Fig.~\ref{fig3} by a solid line and its envelopes (\ref{eq64}) 
by dashed lines.

The wavelength (\ref{eq4}) in the sine-Gordon equation case is given by the expression 
\begin{equation}\label{eq65}
  L=4\frac{\sqrt{1-v^2(e)}}{v(e)}K(e).
\end{equation}
For applicability of Whitham's modulation theory it is necessary, naturally, that the
wavelength $L$ must be much smaller than the size of the whole wave structure. From practical
point of view one can notice that $L$ tends to infinity as $v\to0$ and the right-edge velocity 
$dx_R/dt=v(0)$ (\ref{eq60}) vanishes at $t=-9/2$. Consequently, our particular solution is
only applicable for $t<0,\,|t|\gg5$.

\begin{figure}
\includegraphics[width=8cm]{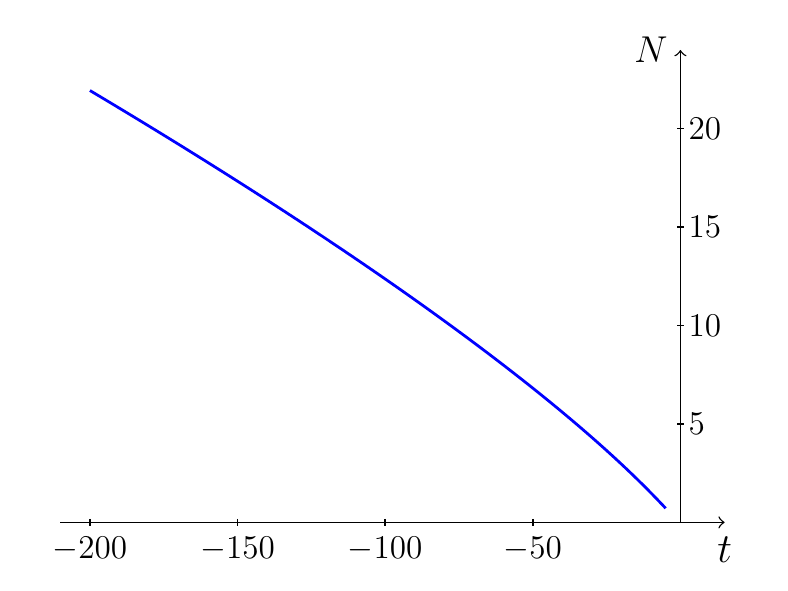}
\caption{Number of oscillations in the nonlinear wave packet as a function of time.
}
\label{fig4}
\end{figure}

Number of oscillations in the nonlinear wave structure is equal approximately to 
\begin{equation}\label{eq66}
  N\cong\int_{x_L}^{x_R}\frac{dx}{L}=\int_0^1\frac{|dx/de|}{L(e)}\,de,
\end{equation}
and, as one can see in Fig.~\ref{fig4}, it decreases with time taking values about unity at $t=-5$.
that is at the boundary of applicability of Whitham's theory. This conclusion about decrease of
number of oscillations agrees with the fact that by virtue of inequality $v<1$ the phase velocity 
$V=1/v(0)$ of the right edge is always greater than the group velocity $v(0)$ of this edge, so that
oscillations leave with time the region of nonlinear oscillations at the small-amplitude edge of
the structure. It is worth noticing that situation here is opposite to the theory of dispersive
shock waves in modulationally stable systems \cite{gp-73} where oscillations enter to the Whitham
region of nonlinear oscillations at the small-amplitude edge \cite{gp-87}. This permits one to
find the number of solitons formed from an intensive initial simple-wave pulse at asymptotically
large times \cite{kamch-21a} (see also \cite{egkkk-07,egs-08,kamch-20,kamch-21b,cdb-k-21}).

We remark that according to Eqs.~(\ref{eq38}) the change of the sign of the function $\mathcal{W}$
is equivalent to the change of signs of $x$ and $t$. Then we get the solution which describes an
expanding with growth of $t$ wave packet similar to formation of the nonlinear wave structure in
evolution of a step-like pulse in the NLS equation theory (see, e.g.,\cite{kamch-92,ehkk-93,bk-94,kamch-97}),
but in this case such an evolution is not self-similar and finding the law of motion of the 
small-amplitude edge is beyond the Whitham-Gurevich-Pitaevskii theory.

Evolution of wave packets corresponding to Legendre functions of the 2nd kind used in
Eq.~(\ref{eq46}) is qualitatively similar and we shall not dwell on details here.

\section{Conclusion}

As is known, the gas dynamics methods can be used for description of evolution of unstable systems 
(see, e.g., \cite{zt-91}). Creation of the Whitham modulation theory \cite{whitham-65,whitham-77}, 
discovery of the inverse scattering transform method \cite{zmnp-80} and development of the general
theory of hydrodynamic type systems \cite{dn-89} permitted one to extend these methods to the vast
area of dispersive shock waves theory (see, e.g., review articles \cite{kamch-21a,eh-16}). Though
these methods are applicable to modulationally unstable systems as well (see, e.g., the papers
\cite{akns-73,takhtajan-74,kk-76,gn-03} on the sine-Gordon equation theory),
their practical applications to one-phase solutions in unstable systems was
limited so far only to step-like situations \cite{kamch-92,ehkk-93,bk-94} (see also \cite{kamch-97} 
and references therein). Evolution of the instability region was discussed in Ref.~\cite{ks-21}
also only for the small-amplitude edge and evolution of modulation parameters in the whole
region of nonlinear oscillations was not considered. In this paper, we have given the exact
solution of the Whitham modulation equations in the sine-Gordon model. Its application to a
concrete example demonstrates quite nontrivial evolution of a nonlinear wave packet when its
contraction is accompanied by going out of waves from the region of nonlinear oscillations.
When the number of oscillations becomes of the order of unity, the Whitham theory becomes
inapplicable and some other methods should be used for description of the wave evolution.

One may hope that this approach will be effective for other versions of the nonlinear
Klein-Gordon equation which describe, in particular, excitation of water waves by wind 
(see \cite{kl-95}) or propagation of electromagnetic waves in nonlinear media
(see, e.g., \cite{su-06,sazonov-14,su-18}).

I am grateful to S.~Yu.~Dobrokhotov, E.~A.~Kuznetsov, and S.~V.~Sazonov for useful
discussions.

\end{document}